\documentclass[conf]{new-aiaa} 
\usepackage[utf8]{inputenc}

\usepackage{graphicx}
\usepackage{amsmath}

\usepackage{amsmath} 

\usepackage{amsfonts}

\usepackage{amsthm}

\theoremstyle{definition}

\usepackage{float} 
\usepackage[skip=0em,font=footnotesize]{caption}

\usepackage[margin = 1in]{geometry}

\usepackage{tikz}
\usetikzlibrary{shapes,arrows,calc,positioning}
\tikzstyle{bigblock} = [draw, fill=blue!20, rectangle, 
    minimum height=6em, minimum width=8em]
\tikzstyle{medblock} = [draw, fill=blue!20, rectangle, 
    minimum height=4em, minimum width=4em]    
\tikzstyle{mux} = [draw, fill=black!20, rectangle, 
    minimum height=5em, minimum width=0.1em]    
\tikzstyle{smallblock} = [draw, fill=blue!20, rectangle, 
    minimum height=2em, minimum width=3em]
    
\tikzstyle{data_block} = [draw, fill=green!20, rectangle, 
    minimum height=2em, minimum width=3em]
\tikzstyle{ops_block} = [draw, fill=blue!20, rectangle, 
    minimum height=2em, minimum width=3em]    
\tikzstyle{est_block} = [draw, fill=red!20, rectangle, 
    minimum height=2em, minimum width=3em]    
    
\tikzstyle{sum} = [draw, fill=blue!20, circle, node distance=1cm,minimum height=0.5cm]
\tikzstyle{signal} = [coordinate]
\tikzstyle{pinstyle} = [pin edge={to-,thin,black}]
\tikzstyle{block} = [draw, fill=blue!20, rectangle, 
    minimum height=3em, minimum width=9em]
\tikzstyle{blockS} = [draw, fill=blue!20, rectangle, 
    minimum height=3em, minimum width=4em]    
\tikzstyle{input} = [coordinate]
\tikzstyle{output} = [coordinate]
\usetikzlibrary{matrix}

\usetikzlibrary{positioning}
\usetikzlibrary{math} 

\tikzstyle{gain} = [draw, fill=blue!20, regular polygon, regular polygon sides=3, shape border rotate=30]
\tikzstyle{gain_vert} = [draw, fill=blue!20, regular polygon, regular polygon sides=3, shape border rotate=0, minimum height=1em]

\usepackage{hyperref}
\usepackage{xcolor}
\hypersetup{
    colorlinks,
    linkcolor={blue!100!black},
    citecolor={blue!50!black},
    urlcolor={blue!80!black}
}

\newcommand{\bc}{\begin{center}}
\newcommand{\ec}{\end{center}}
\newcommand{\benum}{\begin{enumerate}}
\newcommand{\eenum}{\end{enumerate}}
\newcommand{\nn}{\nonumber}
\newcommand{\matl}{\left[ \begin{array}}
\newcommand{\matr}{\end{array} \right]}

\renewcommand{\matl}{\begin{bmatrix}}
\renewcommand{\matr}{\end{bmatrix}}

\newcommand{\matls}{\left[ \begin{smallmatrix}}
\newcommand{\matrs}{\end{smallmatrix} \right]}
\newcommand{\isdef}{\stackrel{\triangle}{=}}

\newcommand{\vect}[1]{\overset{\rightharpoonup}{#1}}

\newcommand{\rmA}{{\rm A}}
\newcommand{\rmB}{{\rm B}}
\newcommand{\rmC}{{\rm C}}

\newcommand{\rmE}{{\rm E}}

\newcommand{\rmM}{{\rm M}}
\newcommand{\rmN}{{\rm N}}

\newcommand{\rmP}{{\rm P}}

\newcommand{\rmX}{{\rm X}}

\newcommand{\rma}{{\rm a}}

\newcommand{\rmc}{{\rm c}}
\newcommand{\rmd}{{\rm d}}

\newcommand{\rmf}{{\rm f}}

\newcommand{\rmn}{{\rm n}}

\newcommand{\rmq}{{\rm q}}

\newcommand{\rms}{{\rm s}}

\newcommand{\BBR}{{\mathbb R}}

\newcommand{\SL}{{\mathcal L}}
\newcommand{\SM}{{\mathcal M}}

\newcommand{\shiftq}{{\textbf{\textrm{q}}}}

\newcommand{\framedot}[2]{\stackrel{{\rm #1}\bullet}{#2}}

\newcommand{\ihat}{ {\hat \imath}}
\newcommand{\jhat}{ {\hat \jmath}}
\newcommand{\khat}{ {\hat k}}

\usepackage{enumitem,amssymb}
\newlist{todolist}{itemize}{2}
\setlist[todolist]{label=$\square$}
\usepackage{pifont}
\usepackage[colorinlistoftodos,textwidth=1 in,textsize=footnotesize]{todonotes}

\title{Thrust Regulation in a Solid Fuel Ramjet using \\ Dynamic Mode Adaptive Control}

\author{
Parham Oveissi\footnote{Graduate Research Assistant, Department of Mechanical Engineering, University of Maryland, Baltimore County, 1000 Hilltop Circle, Baltimore, MD 21250.}, 
Gohar T. Khokhar\footnote{Postdoctoral Research Associate, Department of Aerospace \& Mechanical Engineering, University of Arizona, 1130 N. Mountain Avenue, Tucson, AZ 85721. AIAA Member.},
Kyle Hanquist\footnote{Assistant Professor, Department of Aerospace \& Mechanical Engineering, University of Arizona, 1130 N. Mountain Avenue, Tucson, AZ 85721. AIAA Senior Member.}, 
Ankit Goel\footnote{Assistant Professor, Department of Mechanical Engineering, University of Maryland, Baltimore County, 1000 Hilltop Circle, Baltimore, MD 21250.}
}

\title{Swarm-optimized Adaptive Augmentation of Missile Autopilot}
\author{
Alexander Dorsey\footnote{Undergraduate Student, Department of Mechanical Engineering, 1000 Hilltop Circle, Baltimore, MD 21250.}, 
Parham Oveissi\footnote{Graduate Student, Department of Mechanical Engineering, 1000 Hilltop Circle, Baltimore, MD 21250.}, 
Jeffrey D. Barton\footnote{Principal Staff, Johns Hopkins University Applied Physics Laboratory, Laurel, MD, 20723.},
Ankit Goel\footnote{Assistant Professor, Department of Mechanical Engineering, 1000 Hilltop Circle, Baltimore, MD 21250.}
}

\begin{document}

\maketitle

\begin{abstract}
This paper considers the problem of optimizing a missile autopilot.
In particular, the paper investigates the application of an online learning technique to learn and optimize the gains of a three-loop topology autopilot for a planar missile modeled with nonlinear dynamics and nonlinear aerodynamics forces and moments. 
The classical autopilot for a missile is based on a three-loop topology, where each loop consists of tunable proportional gains. 
An adaptive three-loop autopilot is constructed by augmenting the classical autopilot's fixed-gain controllers with a learning-based controller, which is recursively optimized using retrospective cost optimization. 
Numerical simulations show that online learning improves the tracking performance of the classical autopilot in both nominal and off-nominal interception scenarios.  

\end{abstract}

\section{Introduction}

The design of autopilots for missiles is a challenging control problem due to several factors, including nonlinear dynamics, nonminimum phase behavior due to nose-mounted gyro sensor and tail-fin actuation, 
and uncertain aerodynamic loading. 
Furthermore, in a typical mission, a missile undergoes a range of Mach numbers and aggressive maneuvers, which results in strongly varying aerodynamic loading. 
Model-based control of missiles is thus impractical. 
Instead, a common classical missile autopilot is based on a three-loop topology that uses the IMU data, including the acceleration and angular velocity measurements, with flight-scheduled gains. 
The gain-scheduled controller consists of gains scheduled to several operating conditions. 
However, designing the controller gains for each operating condition is expensive and time-consuming.

Several nonlinear adaptive control techniques have been investigated for missile autopilot design. 
Model-based nonlinear control techniques such as adaptive backstepping is explored in \cite{KIM2004149}, $\SL_1$ control is explored in \cite{L1Adapti69}, MRAC is explored in \cite{Ouda2018}, 
input-output linearization is explored in \cite{TSOURDOS2005373, 6703862}, and 
Lyapunov-based approaches are explored in \cite{Adaptive54, HOU2013741}.
In this paper, we consider the retrospective cost adaptive control (RCAC) to recursively optimize the gains of a classical three-loop autopilot. 
RCAC has been previously investigated for the missile control application. 
In particular, an auto-regressive-moving-average adaptive control law was used to generate pitch rate commands in \cite{7170840}, 
whereas RCAC was used to improve the missile's ability to intercept a target under sensor failures in \cite{Investig5}.
However, this paper investigates the ability of RCAC to tune the gains of the three-loop autopilot. 
In particular, we augment the fixed-gain controllers in the three-loop autopilot with adaptive controllers in parallel. 
This architecture allows RCAC to both improve the autopilot's tracking performance as well as automatically tune the gains of the autopilot from measured data instead of models.

The paper is organized as follows. 
Section \ref{sec:missiledynamics} briefly reviews the nonlinear dynamics of a missile and the classical three-loop autopilot topology.  
Section \ref{sec:3LA_augmented} presents the adaptive augmentation of the three-loop autopilot. 
Three scenarios are presented in Section \ref{sec:numerical} to investigate the performance of the adaptive autopilot in nominal and off-nominal conditions. 
Finally, the paper concludes with a summary in Section \ref{sec:conclusions}.

\section{Missile Dynamics and Control}
\label{sec:missiledynamics}
This section briefly reviews the equations of motion of a missile and the classical three-loop autopilot.

\subsection{Longitudinal Dynamics}
\begin{figure}[h]
    \centering
    \includegraphics[width=0.5\columnwidth]{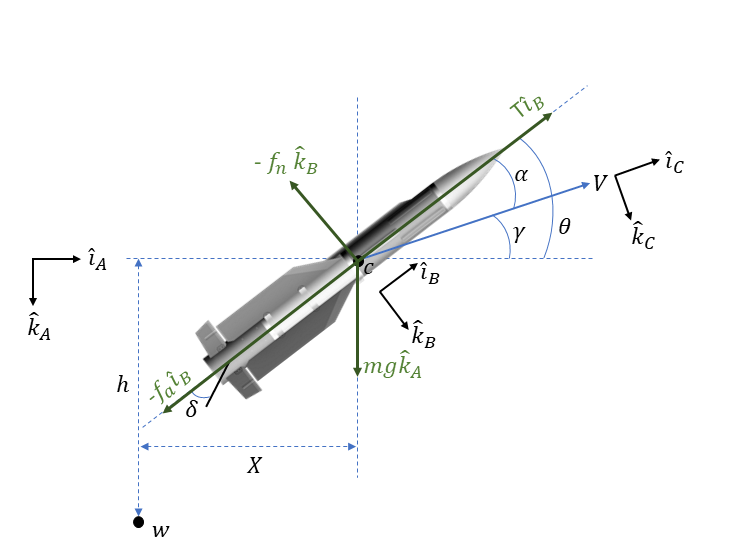}
    \caption{
    Free body diagram of a missile. 
    }
    \label{fig:missile_diagram}
\end{figure}

Let $\rm F_A$ be an inertial frame and let $\rm F_B$ be a frame fixed to the missile body as shown in \autoref{fig:missile_diagram}.
The frame $\rm F_B$ is obtained by rotating it about $\jhat_\rmA$ by the \textit{pitch angle} $\theta,$ 

and thus $\vect \omega_{\rm B/A} = \dot \theta \hat \jmath_\rmB.$
Let $\rmc$ denote the center of mass of the missile and let $w$ denote a point with zero inertial acceleration. Let $\rm F_C$ be a frame such that 
$\ihat_\rmC = \hat v_{\rmc/w/\rmA}$ and 
$\jhat_\rmC=\jhat_{\rm B},$
where $\hat v_{\rmc/w/\rmA}$ is the unit vector along the velocity $\vect v_{\rmc/w/\rmA}$ of the missile. 
Therefore, $\vect v_{\rmc/w/\rmA} = V \ihat_\rmC,$ where $V$ is the magnitude of the missile velocity. 
Note that the frame $\rm F_C$ is obtained by rotating it about $\jhat_\rmA$ by the \textit{flight path angle }$\gamma.$ 

The \textit{angle of attack} is thus $\alpha \isdef \theta - \gamma.$

The equations of motion of a missile are
\begin{align}
    \dot M
        &=
            \frac{T}{ma}  \cos (\alpha)
                - \frac{g}{a} \sin (\gamma)
                - \frac{\rho a M^2 S}{2m} \big[C_\rmN \sin (\alpha) + C_\rmA \cos (\alpha) \big]
        , \label{eq:Mdot_v3}\\
    \dot \gamma
        &=
              \frac{T}{maM} \sin (\alpha)
            - \frac{g}{aM}\cos (\gamma) 
            + \frac{\rho a M S}{2m} \big[C_\rmN \cos (\alpha) - C_\rmA \sin (\alpha) \big]
            , \label{eq:V_gammadot_v3} \\
    \dot \theta 
        &= 
            q,
            \label{eq:thetadot_v2}
    \\
    \dot q
        &=
            \frac{\rho a^2 M^2 S d}{2I_y} C_\rmM
    ,\\
    \dot h 
    &= 
    Ma\sin(\gamma), \\
    \dot X 
    &= 
    Ma\cos(\gamma),
    \label{eq:MD_Xdot}
\end{align}
where 
$m$ is the mass of the missile, 
$a$ is the speed of sound, 
$M$ is the Mach number,
$T$ is the thrust applied to the missile along $\ihat _\rmB, $
$I_y$ is the moment of inertia, 
$q$ is the pitch rate, 
$h$ is the altitude, and 
$X$ is the downrange. 
The derivation of the equations of motion \eqref{eq:Mdot_v3}-\eqref{eq:MD_Xdot} and the vehicle parameters used in this study are detailed in Appendix A. 
The aerodynamic coefficients $C_\rmN, C_\rmA, $ and $C_\rmM$ are nonlinear functions of the angle of attack $\alpha$, fin deflection angle $\delta,$ and the pitch rate $q,$ as shown in \eqref{eq:def_C_N}-\eqref{eq:def_C_M}.

\subsection{Three-loop Autopilot}
\label{sec:3LA}
\begin{figure}[h]
    \centering
    \resizebox{0.7\columnwidth}{!}{ 
    \begin{tikzpicture}[auto, node distance=2cm,>=latex',text centered]
    
        \node at (-3,0) (reference) {$a_{z,\rm ref}$};
        \node [smallblock, fill=green!20, right = 3 em of reference] (Controller) 
        {$\begin{array}{c}{\text{Three-loop} } \\ {\rm Autopilot}\end{array}$};

        \node [smallblock, fill=blue!20, right = 3 em of Controller] (Nonlinear) 
        {$\begin{array}{c}{\rm Actuator} \\ {\rm Dynamics}\end{array}$};

        \node [smallblock, right = 3 em of Nonlinear] (Plant) 
        {$\begin{array}{c}{\rm Missile} \\ {\rm Dynamics}\end{array}$};

        \node [smallblock, right = 3 em of Plant] (IMU) 
        {IMU};
        
        \draw[->] (reference) -- node[xshift = 0em, yshift = .2em]{} (Controller);
        \draw[->] (Controller)-- node[xshift = 0em, yshift = .2em]{$u$} (Nonlinear);
        \draw[->] (Nonlinear)-- node[xshift = 0em, yshift = .2em]{$\delta$} (Plant);
        \draw[->] (Plant) -- node[xshift = 0em, yshift = .2em]{} (IMU) ;
        \draw[->] (IMU.east) node[xshift = 1em, yshift = 0.9em]{$y$} -- +(1,0) --  +(+1,-1.5) 
                    -| (Controller.270);
        
    \end{tikzpicture}
    }
    \caption{Control architecture to track normal acceleration commands. }
    \label{fig:BSL}
\end{figure}
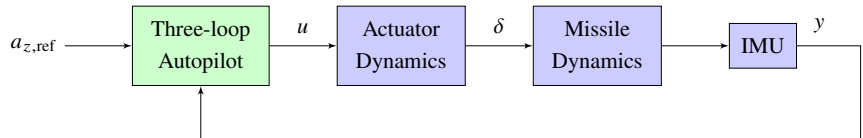

This section briefly reviews the three-loop autopilot. 
Several multi-loop autopilot topologies for normal-acceleration tracking are described in more detail in \cite{mracek2005missile}.
%
%
The control architecture for normal acceleration tracking is shown in Figure \ref{fig:BSL}.
In practice, the normal acceleration reference $a_{z,\rm ref}$ is given by the guidance law.
The three-loop autopilot uses the reference normal acceleration and the measurements of the normal acceleration and the pitch rate to compute the fin deflection command $u$.

\begin{figure}[h]
    \centering
    \resizebox{0.7\columnwidth}{!}
    {
    \begin{tikzpicture}
    
        \node at (-3,0) (reference) {$a_{z,\rm ref}$};

        \node [below = 7 em of reference] (az) {$a_{z}$};
        \node [right = 5 em of az] (q) {$q$};
        \node[sum, right = 2 em of reference] (sum1) {};
        \node[gain_vert, fill=green!20, below = 2 em of sum1] (Kss) {$K_{\rma_z}$};
        
        \node[gain, fill=green!20, right = 2 em of sum1] (Ka) {$K_{\rm a}$};
        \node[sum, right = 2 em of Ka] (sum2) {};
        \node[smallblock, right = 2 em of sum2] (Integrator) {$\int$};
        
        \node[sum, right = 2 em of Integrator] (sum3) {};

        \node[gain_vert, fill=green!20, below = 1.5 em of sum2] (Ktheta) {$K_{\theta}$};
        \node[gain_vert, fill=green!20, below = 1.5 em of sum3] (Kq) {$K_{\rm q}$};
        
        \draw[->] (reference) -- node[xshift = 0em, yshift = .2em]{} (sum1);
        \draw[->] (sum1) -- (Ka);
        \draw[->] (Ka) -- (sum2);
        \draw[->] (sum2)-- (Integrator);
        \draw[->] (Integrator) -- (sum3);
        \draw[->] (sum3) -- +(1,0) node[xshift = 1em]{$u$};
        \draw[->] (az) -| (Kss);
        \draw[->] (Kss) -- (sum1.270) node[xshift = -0.5em, yshift = -0.5em]{$-$};
        \draw[->] (q) -| (Ktheta.south);
        \draw[->] (Ktheta.90) -| (sum2.270) node[xshift = -0.5em, yshift = -0.5em]{$+$};
        \draw[->] (q) -| (Kq.south);
        \draw[->] (Kq.90) -| (sum3.270) node[xshift = -0.5em, yshift = -0.5em]{$+$};

    \end{tikzpicture}
    }
    \caption{Classical three loop controller topology \cite{mracek2005missile}. }
    \label{fig:3LA_diagram}
\end{figure}
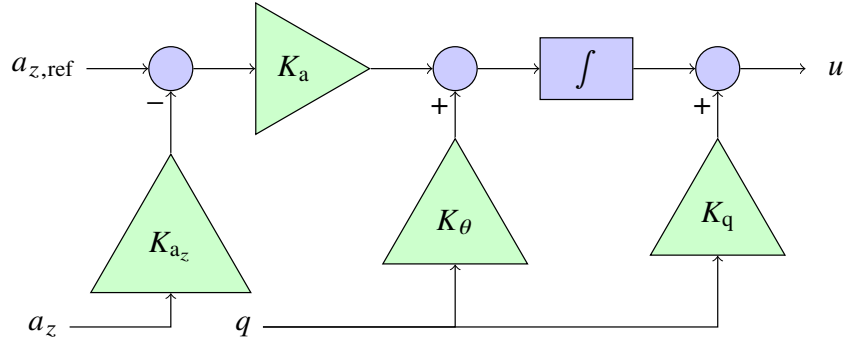

In particular, the control signal $u$ generated by the three-loop autopilot is 
\begin{align}
    u
        = 
            K_q q +
            \int ( K_\theta q + K_\rma(a_{z, \rm ref} - K_{\rma_z} a_z) 
            ) \rmd t, 
    \label{eq:u_3LA}
\end{align}
where
$K_\rmq, K_\theta, K_\rma,$ and $K_{\rma_z}$ are the tunable proportional gains. 
In this work, as considered in \cite{mracek2005missile},  we set 
$K_\rmq = 0.464, K_\theta = 15.62474, K_\rma = 0.2446459,$ and $K_{\rma_z} = 0.9278,$ $a_{z, \rm ref}$ is the normal acceleration command, $a_{z}$ is the sensed acceleration at the IMU given by
\begin{align}
    a_z
        \isdef
            a_{z, \rm CG} - \dot q d_{\rm IMU},
\end{align}
where $d_{\rm IMU}=0.5 \ \rm m$ is the distance from the center of gravity to the IMU, and the gravity-corrected normal acceleration of the center of gravity  $a_{z, \rm CG}$ is
\begin{align}
    a_{z, \rm CG}
        =
            \dfrac{\rho a^2 M^2 S}{2{m}} C_\rmN. 
\end{align}
The IMU provides the output vector $y$ that contains the measurements of the normal acceleration $a_z$ and pitch rate $q.$
The implementation of the control law \eqref{eq:u_3LA} is shown in Figure \ref{fig:3LA_diagram}.
In the rest of the paper, the three-loop autopilot with fixed gains is denoted by F-TLA.

\section{Adaptive TLA }
\label{sec:3LA_augmented}
\begin{figure}[h]
    \centering
    \resizebox{0.7\columnwidth}{!}
    {
    \begin{tikzpicture}
    [auto, node distance=2cm,>=latex',text centered, font=\sffamily\footnotesize]
        
        \draw[draw=black, fill=red!5] (-1.4,0.8) rectangle ++(5.7,1.8) node[xshift = -5.5em, yshift = -0.8em]{\textbf{Adaptive Augmentation}};
        \node at (-3,0) (reference) {$a_{z,\rm ref}$};

        \node [below = 7 em of reference] (az) {$a_{z}$};
        \node [right = 5 em of az] (q) {$q$};
        \node[sum, right = 2 em of reference] (sum1) {};
        \node[gain_vert, fill=green!20, below = 2 em of sum1] (Kss) {$K_{\rma_z}$};

        \node[gain, fill=green!20, right = 2 em of sum1] (Ka) {$K_{\rm a}$};
        
        \node[smallblock, fill=red!20] at (1.5,1.5) (AdaptivePI) 
        {$\begin{array}{c}{\text{Adaptive} } \\ {\text{Controller}}\end{array}$};
        \node[smallblock, fill=red!20, right = 2 em of AdaptivePI] (zoh) {ZOH};
        \node[smallblock, fill=red!20, left = 2 em of AdaptivePI] (A2D) {A2D};
        
        \node[sum, right = 2 em of Ka] (sum2) {};
        \node[smallblock, right = 2 em of sum2] (Integrator) {$\int$};
        
        \node[sum, right = 2 em of Integrator] (sum3) {};

        \node[gain_vert, fill=green!20, below = 1.5 em of sum2] (Ktheta) {$K_{\theta}$};
        \node[gain_vert, fill=green!20, below = 1.5 em of sum3] (Kq) {$K_{\rm q}$};
        
        \draw[->] (reference) -- node[xshift = 0em, yshift = .2em]{} (sum1);
        \draw[->] (sum1) -- (Ka);
        \draw[->] (Ka) -- (sum2);
        \draw[->] (sum2)-- (Integrator);
        \draw[->] (Integrator) -- (sum3);
        \draw[->] (sum3) -- +(1,0) node[xshift = 1em]{$u$};
        \draw[->] (az) -| (Kss);
        \draw[->] (Kss) -- (sum1.270) node[xshift = -0.5em, yshift = -0.5em]{$-$};
        \draw[->] (q) -| (Ktheta.south);
        \draw[->] (Ktheta.90) -| (sum2.270) node[xshift = -0.5em, yshift = -0.5em]{$+$};
        \draw[->] (q) -| (Kq.south);
        \draw[->] (Kq.90) -| (sum3.270) node[xshift = -0.5em, yshift = -0.5em]{$+$};

        \draw[->] (sum1.north) |- (A2D.180);
        \draw[->] (A2D.0) -- (AdaptivePI.west);
        \draw[->] (AdaptivePI.0) -- node[xshift = 0em, yshift = 0em]{$u_k$} (zoh.180);
        \draw[->] (zoh) -| node[xshift = 0em, yshift = 0em]{$u_\rma$} (sum3.90);
    \end{tikzpicture}
    }
    \caption{Adaptive three loop controller topology. }
    \label{fig:3LA_internal}
\end{figure}
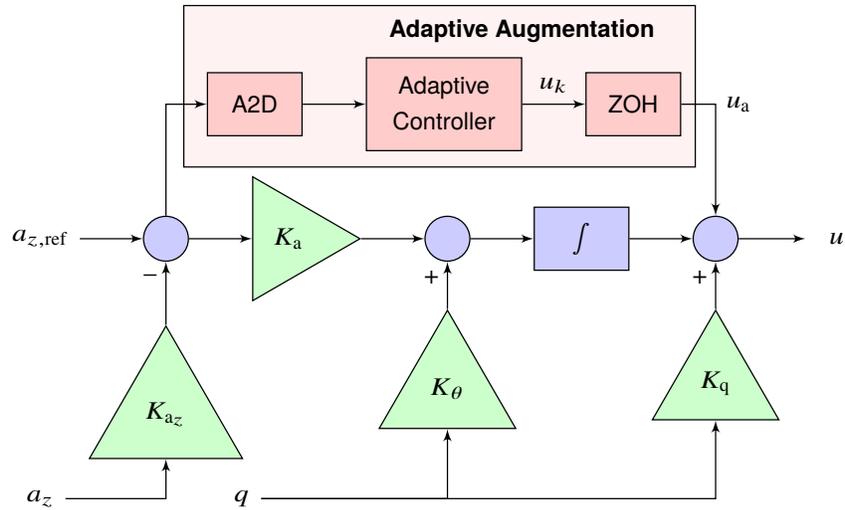

In the adaptive augmentaion, shown in Figure \ref{fig:3LA_internal}, the input signal computed by the three-loop autopilot is modified as
\begin{align}
    u(t) = u_{\rm TLA }(t) + u_{\rm a}(t),
\end{align}
where $u_{\rm TLA }(t)$ is the control signal computed using \eqref{eq:u_3LA} and $u_\rma(t)$ is the control signal computed by the adaptive controller. 
In particular, the control signal $u_\rma(t)$ is computed by the RCAC algorithm. 
Since RCAC is a discrete-time algorithm that updates the control signal at a fixed timestep, the control $u_\rma(t)$ in between the control updates is held constant. 
Letting $T_\rms$ denote the timestep, for $t \in (k T_\rms, (k+1) T_\rms),$ the control $u_\rma(t)$ is thus given by
\begin{align}
    u_\rma(t) 
        =
            u_{k},
\end{align}
where $k$ is the iteration number and $u_k$ is the control signal computed by RCAC.
In particular, an adaptive dense structured controller, described in \cite{SparceRCAC}, is chosen so that the regressor matrix is given by 
\begin{align}
    {\phi_k} 
        = 
            \begin{bmatrix} 
                u_{k-1} &
                \cdots &
                u_{k-n_c}& 
                z_{k-1}&
                \cdots &
                z_{k-n_c} 
                \gamma_k
            \end{bmatrix},
\end{align}
where
$z_k \isdef a_{z,\rm ref} - a_z,$
$\gamma_k \isdef \sum_{i=1}^k z_i$ is the accumulated error, 
and the adaptive controller gain 
$\theta_k \in \BBR^{2 n_c+1}$ is updated by retrospective cost optimization. 
The adaptive control signal  $u_k$ is thus an adaptive weighting of the regressor matrix, in the form of $u_k =  \Phi_k \theta_k.$ The adaptive weight $\theta_k$ is determined at each time step $k$ by minimizing the retrospective cost function
\begin{align}
    J(k)
        &\isdef
            \sum_{i=1}^{k}
            \lambda^{k-i}
            [\hat{z}^{\mathrm{T}}_{i}R_z\hat{z}_{i}+(\phi_{i}\hat{\theta})^{\mathrm{T}}R_{u}(\phi_{i}\hat{\theta})]
            + \lambda^{k}(\hat{\theta}-\theta _{0})^{\mathrm{T}}R_{\theta}(\hat{\theta}-\theta_{0}),
\end{align}
where 
$\hat z_i \isdef z_i + \phi_{\rmf,i} \hat \theta - u_{\rmf, i} $ is the retrospective performance, 
$\phi_{\rmf, i} \isdef G_\rmf(\shiftq) \phi_i$ and 
$u_{\rmf, i} \isdef G_\rmf(\shiftq) u_i$ are the filtered regressor and the input, 
$\shiftq$ is the forward-shift operator, 
$\lambda$ is a forgetting factor, $R_z$, $R_u$, and $R_\theta$ are hyperparameters, and $G_f$ is a filter.
The RCAC algorithm to compute $u_k$ is described in \cite{rezaPID,oveissi2023learning,oveissi2024adaptive}.

As shown in \cite{rahman2017retrospective}, the cost function that is retrospectively minimized in RCAC includes a filter, $G_\rmf$, which should consist of any nonminimum phase zeros of the system. 
In this work, we update the filter $G_\rmf$ with the real nonminimum phase zeros of the linearized missile dynamics \eqref{eq:Mdot_v3}-\eqref{eq:MD_Xdot}, computed at the current state. 
Thus, the filter $G_\rmf$ is time varying and is updated at every step.
In the rest of the paper, the three-loop autopilot with the adaptive augmentation is denoted by A-TLA.

\section{Numerical Simulation}
\label{sec:numerical}

This section investigates the performance improvements with the adaptive augmentation of the TLA. 
In particular, the performance of the A-TLA is investigated in the case of an arbitrary normal acceleration command and an interception mission scenario, described in Appendix B. 
The missile dynamics and the TLA described in previous sections are simulated with Matlab's ode45 routine. 
In all scenarios considered in this paper, the initial Mach number of the missile is 2.5,
the initial flight path angle is $45$ degrees, 
the initial pitch angle is $45$ degrees, 
the initial pitch rate is $0$ $\rm rad/s,$ 
the initial altitude is $3,500$ $\rm m,$ a constant thrust of $3800$ newtons, and 
the fin deflection angle and its rate are $0$ degrees and $0$ $\rm rad/s,$ respectively.  

\subsection{RCAC Hyperparameter Optimization}

The RCAC hyperparameter $R_z$ is held constant at unity, while $R_u$ and $ R_\theta $ are numerically optimized using a \textit{particle swarm optimziation} (PSO) framework. In a PSO, the parameter vector to optimize is treated as a particle with a position and velocity. A swarm of particles are evaluated and each particle's motion is influenced by the best local and global parameters previously found. The objective of PSO is thus to find the set of parameters that minimize a cost function. 
The cost function used to tune the RCAC hyperparameters in the PSO framework is
\begin{align}
    J_{\rm PSO}
        =
            \frac{1}{N}
            \sum_{i=1}^N
            \frac{2}{g}|z_i|
            + 
            0.2 {\rm max}({|\dot{q_i}|-\dot q_{\rm max}},0),
\end{align}
where $\dot q_{\rm max} = 3$ $\rm degrees/s^2,$
where $N$ is the length of the simulation, and the data $z_i$ and $u_i$ are generated by the closed-loop response to the command  
$a_{z, \rm ref} = 0.5g -8g {\rm sign}(\sin(0.3t)).$

The controller optimized by RCAC is a $ 4$th-order ARMA controller with a built-in integrator. 
Various parameterizations that can be used in the RCAC framework are described in \cite{goel_2020_sparse_para}. 
Note that, in our preliminary work, the cost function is chosen by trial and error. 
In the particle swarm, 
$R_u$ is restricted between $0$ and $20$ and 
$R_\theta$ is restricted between $10^0$ and $10^{15}.$
The RCAC hyperparameters after swarm optimization are $(R_u, R_\theta) = (0.25427, 10^{14.398}).$
In all numerical experiments described next in the paper, the RCAC hyperparameters are fixed. 

\begin{figure}[h]
    \centering
    \includegraphics[width=0.5\columnwidth]{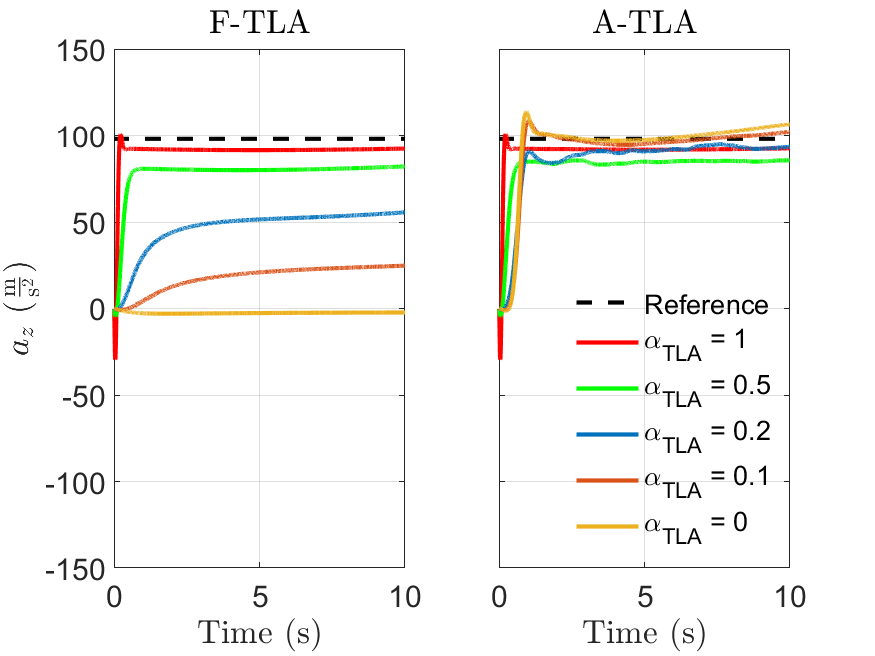}
    \caption{The evolution of the normal acceleration for the F-TLA and A-TLA to a 10 g step command at various scalings of the gains in the TLA .}
    \label{fig:1by2_RCAC_100step}
\end{figure}

Next, the performance improvements due to the adaptive augmentation are investigated. 
The acceleration reference is a constant step of magnitude $10g.$
To demonstrate the performance recovery with the adaptive augmentation, we manually degrade the F-TLA by scaling all of its gains by $\alpha_{\rm TLA} \in \BBR$ 
Figure \ref{fig:1by2_RCAC_100step} shows the closed-loop step response with the F-TLA and the A-TLA for various scaling factor values $\alpha_{\rm TLA}.$
Note that the performance of the F-TLA and the A-TLA is similar in the case where $\alpha_{\rm TLA} = 1.$
On the other hand, the degraded F-TLA's step response performance degrades, as expected.
However, the A-TLA performs similarly for a wide range of $\alpha_{\rm TLA},$ thus demonstrating the performance improvements due to the adaptive augmentation.

\subsection{Harmonic Response}

\begin{figure}[h]
    \centering
    \includegraphics[width=0.5\columnwidth]{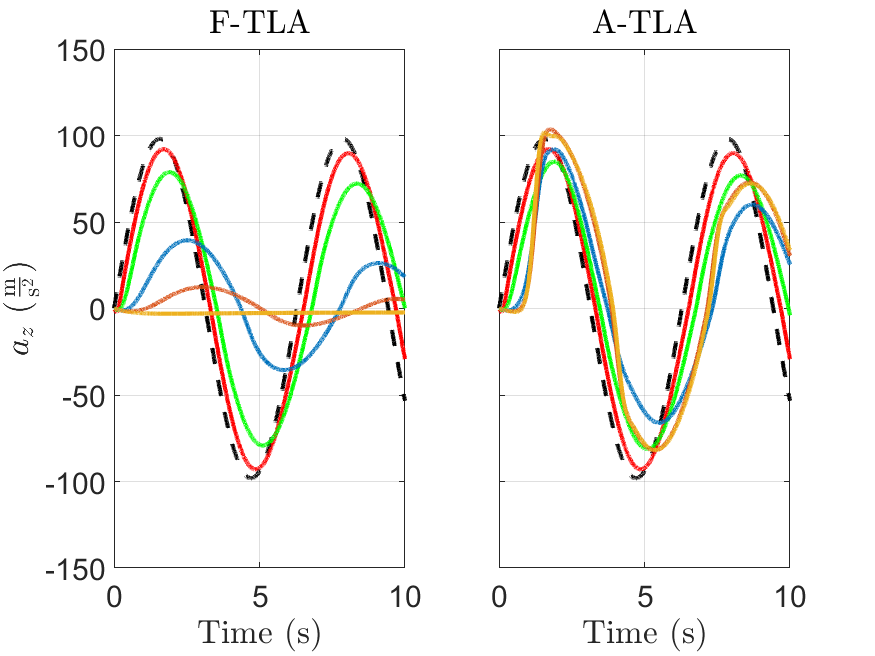}
    \caption{The evolution of the normal acceleration for the F-TLA and A-TLA to a 10 g sinusoidal command.}
    \label{fig:1by2_RCAC_100sin}
\end{figure}

Next, we set $a_{z, \rm ref} = 10 g \sin(t).$ 
Figure \ref{fig:1by2_RCAC_100sin} shows the normal acceleration tracking responsE with the F-TLA and the A-TLA, using the same legend and $\alpha_{\rm TLA}$ scalings decribed previously.

\subsection{Interception Response}

Next, we consider a missile mission where the objective of the missile is to intercept a target. 
To achieve interception, the normal acceleration commands are generated using a proportional guidance law based on the pursuit-evasion dynamics described in Appendix B. 
We consider the evader model described in \cite{islam2022minimum} as the target.
The target's initial speed is 0.85 Mach, its flight path angle is 15 degrees, with an altitude of 4 km, and 1 km of horizontal distance from the missile. 

\begin{figure}[h]
    \centering
    \includegraphics[width=0.5\columnwidth]{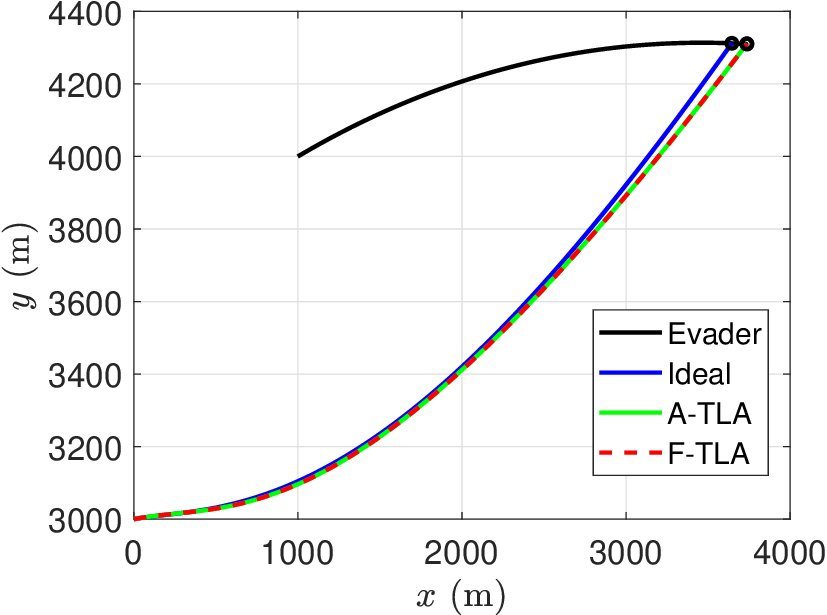}
    \caption{ Interception trajectories with the nominally tuned F-TLA and A-TLA.
    }
    \label{fig:TrajCurvedPath}
\end{figure}

The missile thrust is assumed to be 
\begin{align}
    T(t)
        =
            \begin{cases}
                15,000 \rmN, & t \in (0, 10), \\
                2,000 \rmN, & t \in (10, 20), \\
                0 \rmN, & t > 20.
            \end{cases}
\end{align}
The missile's initial speed is Mach 0.5, its initial flight path angle is 0 degrees, and its initial altitude is 3 km.
The hyperparameters used in the RCAC are the same as those used in previous numerical simulations.
Figure \ref{fig:TrajCurvedPath} shows 
the trajectory of the evader (in solid black), 
the trajectory of an ideal pursuer (in solid blue), 
the trajectory of the missile with the A-TLA (in solid green), and
the trajectory of the missile with the nominally tuned F-TLA (in dashed red).
Note that the ideal pursuer is assumed to be a point mass whose normal acceleration is instantaneously equal to the normal acceleration commanded by the guidance law. 
Furthermore, note that, in the case of a nominally tuned F-TLA, the performance of the A-TLA is similar to that of the F-TLA. 

\begin{figure}[h]
    \centering
    \includegraphics[width=0.5\columnwidth]{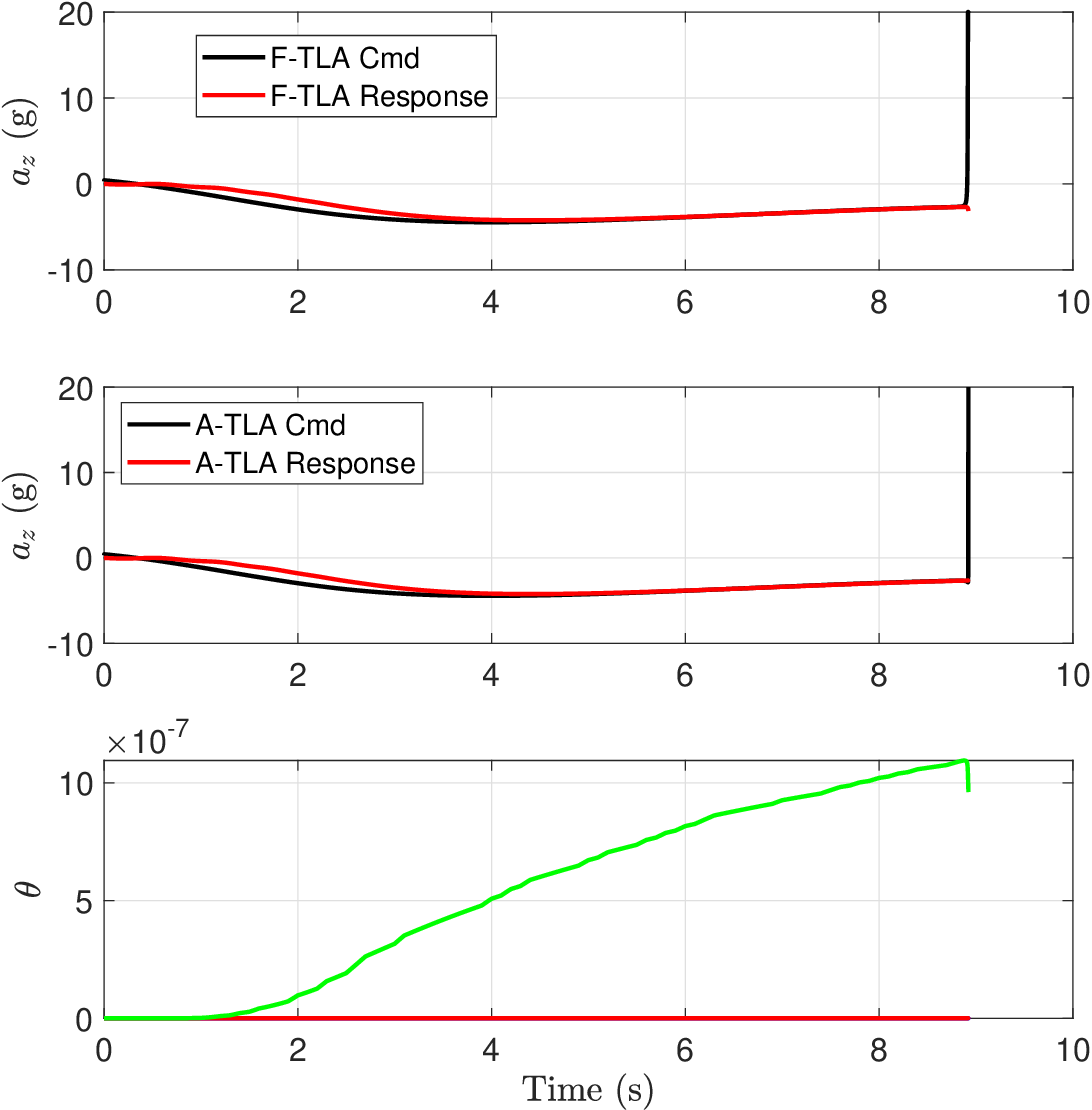}
    \caption{
    Normal acceleration command and response with the nominally tuned F-TLA and A-TLA.
    The third subfigure shows the gains updated by RCAC.
    }
    \label{fig:AccelCurvedPath}
\end{figure}

Figure \ref{fig:AccelCurvedPath} shows the commanded normal acceleration, the response of the missile with the F-TLA and A-TLA. and the gains of the adaptive controller updated by the RCAC algorithm. 
Note that, in the case of a nominally tuned F-TLA, the normal acceleration response of the A-TLA is similar to that of the F-TLA. 
Furthermore, note that the large increase in the normal acceleration command is due to the missile's terminal behaviour. 

\begin{figure}[h]
    \centering
    \includegraphics[width=0.5\columnwidth]{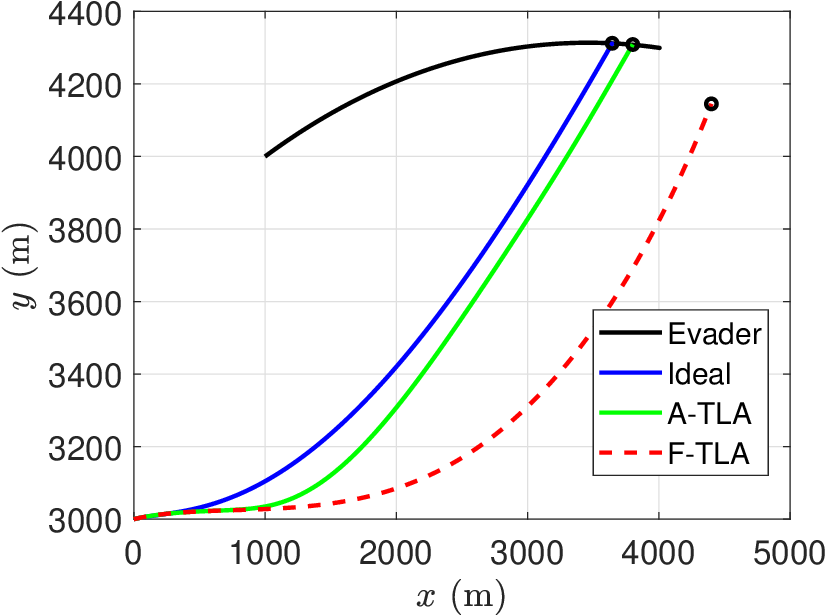}
    \caption{
    Interception trajectories with the off-nominal F-TLA and A-TLA.
    The F-TLA is degraded by scaling the nominal fixed gains by a scalar factor $\alpha{\rm TLA} = 0.2.$
    }
    \label{fig:TrajCurvedPathAlpha02}
\end{figure}

Next, to investigate the performance recovery due to the adaptive augmentation, we degrade the F-TLA by scaling all of the F-TLA gains by the scaler $\alpha_{\rm TLA} = 0.2.$
Figure \ref{fig:TrajCurvedPathAlpha02} shows 
the trajectory of the evader (in solid black), 
the trajectory of an ideal pursuer (in solid blue), 
the trajectory of the missile with the A-TLA (in solid green), and
the trajectory of the missile with the nominally tuned F-TLA (in dashed red).
Note that, in the case of off-nominal F-TLA, the interception performance degrades substantially, however, A-TLA compensates for the degraded F-TLA and recovers the performance.

\begin{figure}[h]
    \centering
    \includegraphics[width=0.5\columnwidth]{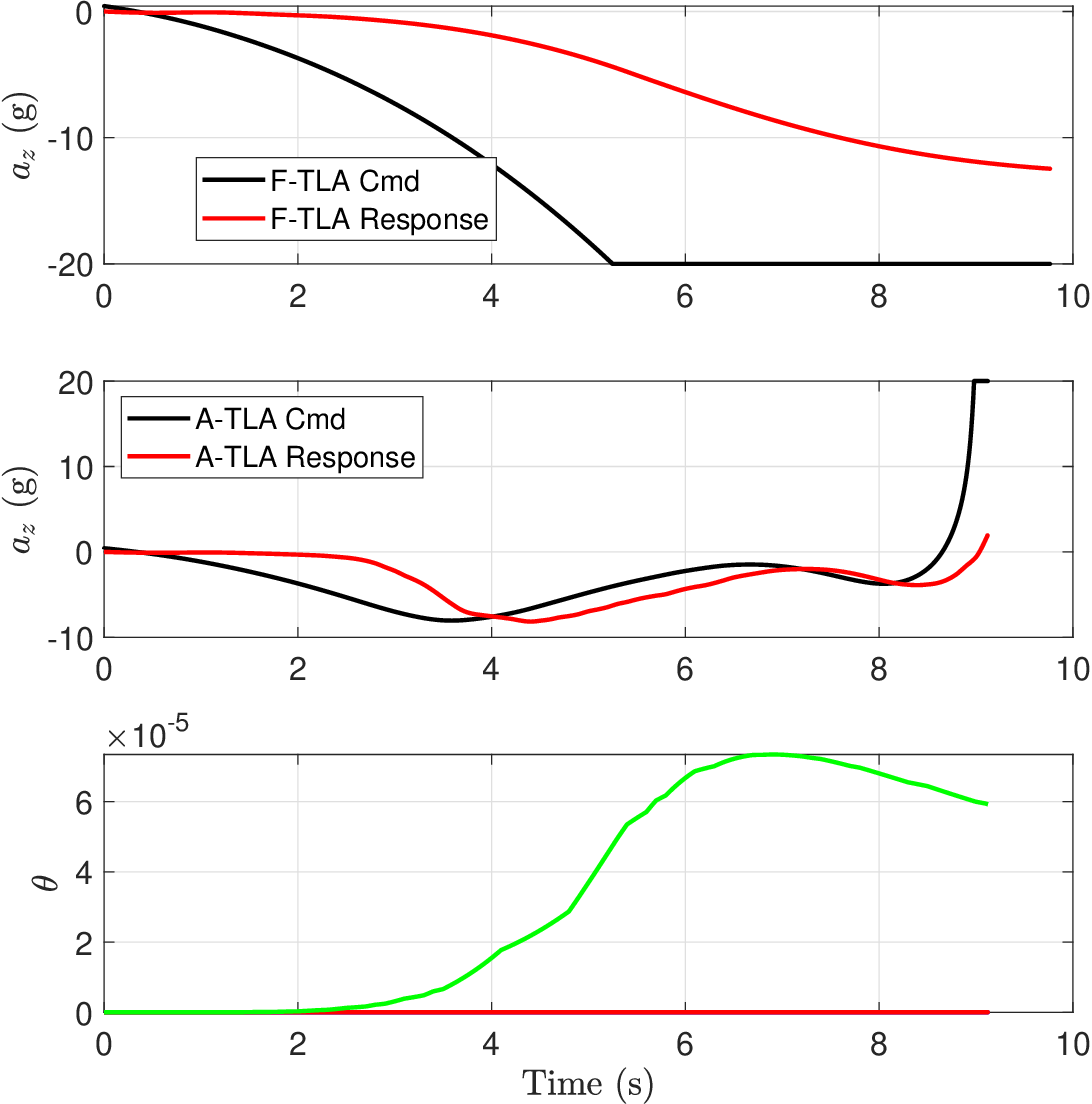}
    \caption{
    Normal acceleration command and response with the off-nominal F-TLA and A-TLA.
    Note that the A-TLA adjusts the controller gains, shown in the third subfigure, to maintain performance. 
    }
    \label{fig:AccelCurvedPathAlpha02}
\end{figure}

Figure \ref{fig:AccelCurvedPathAlpha02} shows the commanded normal acceleration and the response of the missile with the F-TLA and A-TLA. 
Note that, in the case of off-nominal F-TLA, the normal acceleration response degrades. 
However, A-TLA compensates for the degraded F-TLA and generates the required normal acceleration to recover performance.

\begin{figure}[h]
    \centering
    \includegraphics[width=0.5\columnwidth]{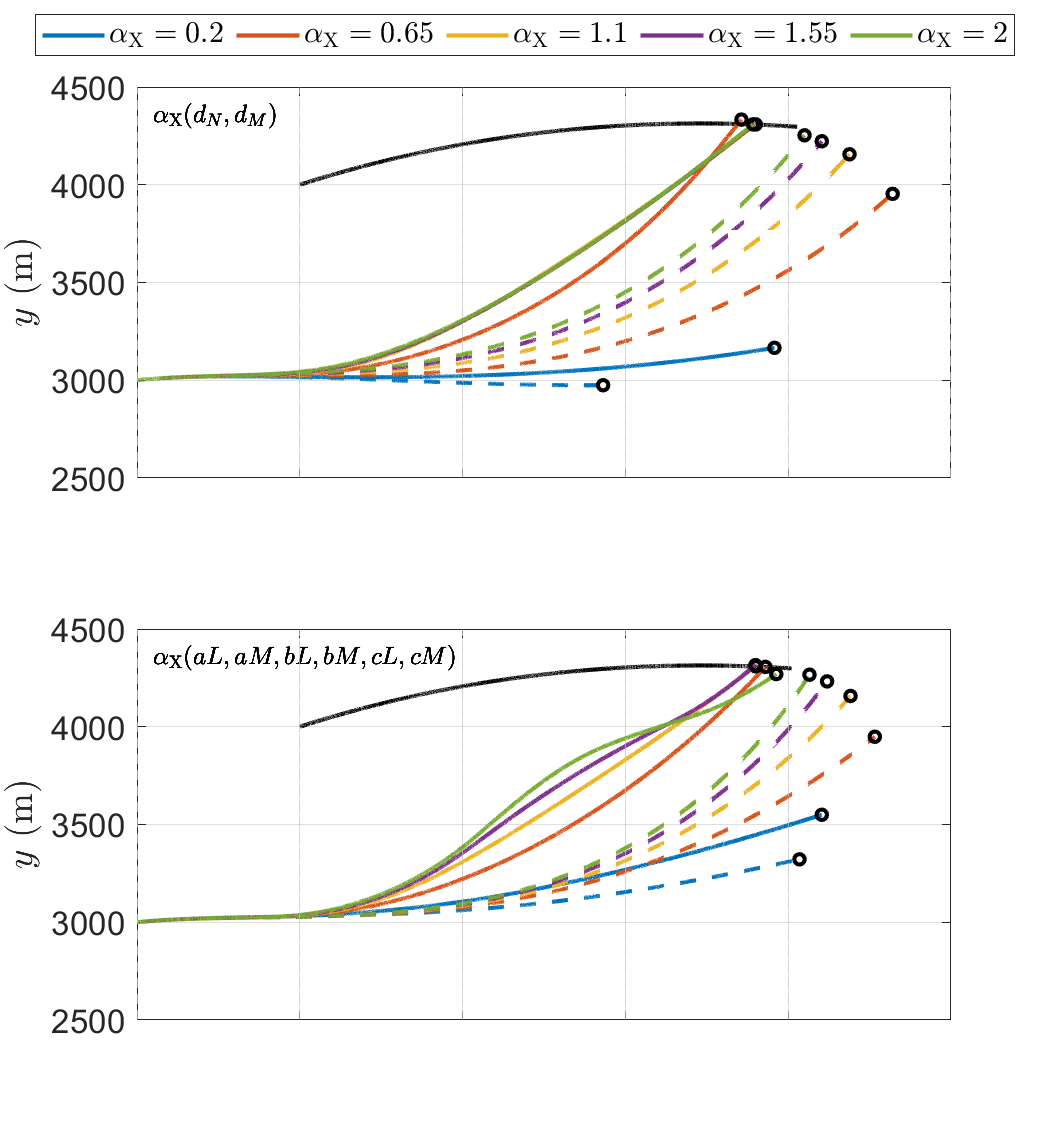}
    \caption{
    Interception trajectories with the F-TLA (dashed lines) and A-TLA (solid lines) in the scenarios where 
    a) the fin-deflection coefficient is scaled, 
    b) the anlge of attack coefficients are scaled,
    Note that the A-TLA maintains performance close to the ideal pursuer for a wider range of parameters. 
    }
    \label{fig:2by2SensitivityStudy}
\end{figure}

Next, to investigate the robustness of the A-TLA, we vary various parameters of the nonlinear missile dynamics. 
In particular, we consider two scenarios, where first, we vary 
the fin-deflection coefficients $d_\rmN$ in \eqref{eq:def_C_N} and $d_\rmM$ in \eqref{eq:def_C_M}, 
and second, we vary all the coefficients which multiply the angle of attack in \eqref{eq:def_C_N} and \eqref{eq:def_C_M}.
In each case, the coefficient to be scaled is multiplied by the scalar factor $\alpha_\rmX.$
In this work, we set $\alpha_\rmX$ to five equispaced values between $0.2$ and $2.$
The interception scenario described in the previous section is considered to investigate the robustness of the the A-TLA. 
Figure \ref{fig:2by2SensitivityStudy} shows the trajectory tracking response of the missile with the F-TLA in dashed lines and with the A-TLA in solid lines.
Note that in each case, the A-TLA compensates for the variation in the missile dynamics and maintains the performance close to the ideal pursuer trajectory, whereas, the performance of the F-TLA degrades substantially in terms of time of flight and miss distance.

\section{Conclusions}
\label{sec:conclusions}
This paper investigated the application of a data-driven output feedback adaptive controller to improve the performance of the classical three-loop autopilot in off-nominal conditions. 
In particular, an adaptive proportional-integral controller, optimized by the retrospective cost adaptive control algorithm, augments a nominally tuned, fixed-gain three-loop autopilot. 
The particle swarm optimization framework was used to optimize the hyperparameters of the adaptive control algorithm. 
The particle-swarm optimized adaptive algorithm was then used to numerically investigate the performance of the adaptive three-loop autopilot in the nominal and off-nominal scenarios. 
Numerical simulations showed that adaptive augmentation maintains the missile's performance close to the ideal pursuer's trajectory in off-nominal scenarios, where the performance with the fixed-gain three-loop autopilot significantly degrades. 

\clearpage
\bibliography{Refs}

\section*{Appendix A: Missile Longitudinal Dynamics}

As shown in Figure \ref{fig:missile_diagram}, $\vect v_{\rmc/w/\rmA} = V \ihat_\rmC,$ and thus
\begin{align}
    \vect a_{\rmc/w/\rmA} 
        &=
            \dot V \ihat_\rmC
            +
            V \framedot{A}{\ihat_\rmC}
        =
            \dot V \ihat_\rmC
            -
            V \dot \gamma \khat_\rmC.
\end{align}
The total force on the missile is
\begin{align}
    \vect f_{\rm B}
        &=
            m g \khat_{\rmA}
            - f_\rmn \khat_{\rmB}
            - f_\rma \ihat_{\rmB}
            + T \ihat_{\rm B}
        \nn \\
        &=
            \big(
                - m g\sin (\gamma) 
                - f_\rmn\sin (\alpha)
                + (T-f_\rma) \cos (\alpha)
            \big)
            \ihat_\rmC 
            + 
            \big(
                m g\cos (\gamma) 
                -f_\rmn \cos (\alpha)
                - (T-f_\rma) \sin (\alpha)
            \big)
            \khat_\rmC,
\end{align}
where $m$ is the mass of the missile,
$g$ is the acceleration due to gravity,
$f_\rmn$ and $f_\rma$ are the normal and axial aerodynamic forces on the missile, and $T$ is the thrust, and 
the total moment relative to $\rmc$ is
\begin{align}
    \vect M_{\rm B/c}
        =
            \SM \jhat_\rmB,
\end{align}
where the aerodynamic forces and moments are parameterized by
\begin{align}
    f_\rmn
        &=
            \frac{1}{2} \rho S V^2 C_\rmN, 
        \quad
    f_\rma
        =
            \frac{1}{2} \rho S V^2 C_\rmA, 
        \quad
    \SM
        =
            \frac{1}{2} \rho S V^2 d C_\rmM, \label{eq:aero_eq}
\end{align}
where $d$ is the chord length, $S$ is the reference surface area, and  $\rho$ is the air density calculated by the International Standard Atmosphere model. 
Furthermore, the aerodynamic coefficients $ C_\rmN,  C_\rmA, $ and $ C_{\rmM} $ are parameterized by
\begin{align}
    C_\rmN 
        &=
            a_\rmN \alpha^3 + b_\rmN\alpha|\alpha| + c_\rmN(2 - {M}/{3})\alpha + d_\rmN\delta,
    \label{eq:def_C_N}
        \\
    C_\rmA 
        &=
            a_\rmA ,
    \label{eq:def_C_A}
        \\
    C_{\rmM} 
        &=
            a_\rmM \alpha^3 +b_\rmM \alpha|\alpha| + c_\rmM({8M}/{3} - 7)\alpha 
            + d_\rmM\delta + e_\rmM q,
    \label{eq:def_C_M}
\end{align} 
where $\delta$ is the fin deflection angle.
The transfer function from the fin deflection angle command $u$ to the fin deflection angle $\delta$ is 
\begin{align}
    G_{\delta u}(s)
        =
            \frac{\omega_a^2}{s^2 + 2 \zeta \omega_a + \omega_a^2}.
\end{align}
where $\zeta$ is the damping ratio and $\omega_a $ is the natural frequency of the actuator. 

Resolving the force $\vect f_{\rm B}$ and the inertial acceleration $\framedot{A}{\vect v}_{\rmc/w/\rmA}$ in $\rm F_{C}$, and using the Newton's second law, $m \framedot{A}{\vect v}_{\rmc/w/\rmA} = \vect f_{\rm B},$ yields
\begin{align}
    m \dot V
        &=
            - m g\sin (\gamma) 
            - f_\rmn \sin (\alpha)
            + (T-f_\rma) \cos (\alpha)
        , \label{eq:Vdot}\\
    -mV \dot \gamma
        &=
            m g\cos (\gamma) 
            -f_\rmn \cos (\alpha)
            - (T-f_\rma) \sin (\alpha). \label{eq:gammadot}
\end{align}
Similarly, resolving the moment $\vect M_{\rm B/c}$ in $\rm F_{C}$ and using the Euler's equation yields
\begin{align}
    I_y \ddot \theta  = \SM. 
    \label{eq: moment qdot}
\end{align}
which can be rewritten in the state-space form as
\begin{align}
    \dot \theta &= q, \quad
    \dot q = \frac{\rho V^2 S d}{2I_y} C_\rmM,
    \label{qdot eq}
\end{align}
where $q$ is the pitch rate. 

For the simulations presented in this work, the values of the parameters parameterizing the aerodynamic coefficient are given in Table \ref{tab:parameter}, and the physical properties of the missile are given in Table \ref{tab:PhysicalParameter}.
\begin{table}[H]
    \centering
    \begin{tabular}{|c|c||c|c|}
        \hline
        Parameter & Value  & Parameter & Value  \\
        \hline
        \hline
        $a_\rmN$ &-19.373 &
        $a_\rmM$ & 40.440 
        \\
        \hline
        $b_\rmN$ & 31.023 &
        $b_\rmM$ & -64.015
        \\
        \hline
        $c_\rmN$ &9.717  &
        $c_\rmM$ &2.922
        \\
        \hline
        $d_\rmN$ & 1.948 &
        $d_\rmM$ & -11.803
        \\
        \hline
        $a_A $ & 0.3005 &
        $e_m$ & -1.719 
        \\
        \hline
        $\omega_a$ & $150$ $\rm rad/s$  &
        $\zeta$ & 0.7
        \\
        \hline
    \end{tabular}
    \caption{Parameter values. }
    \label{tab:parameter}
\end{table}

\begin{table}[H]
    \centering
    \begin{tabular}{|c|c||c|c|}
        \hline
        Parameter & Value  & Parameter & Value  \\
        \hline
        \hline
        Mass $m$ &204.0227 \rm kg &
        $S$ & $0.0409$  $\rm m^2$ 
        \\
        \hline
        $d$ & $0.2286$ $\rm m$ &
        $ I_y$ & $247.4336$ $\rm kg \ m^2$
        \\
        \hline
        $d_{\rm imu}$ & $0.5$ m  &
         & 
         \\
        \hline
    \end{tabular}
    \caption{Physical properties of the missile. }
    \label{tab:PhysicalParameter}
\end{table}

\section*{Appendix B: Interception Guidance}

\begin{figure}[h]
    \centering    
    {
    \begin{tikzpicture}

    \node at (0,0) (P) {};    
    \draw [fill=black] (P) circle [radius=0.050];

    \node at (3,4) (E) {};    
    \draw [fill=black] (E) circle [radius=0.050];

    \draw[->] (P.center) 
            node[xshift=0, yshift=-10] {$\rmP$} circle [radius=0.050]
            -- 
            (E.center) node[xshift=0, yshift=-10] {$\rmE$} circle [radius=0.050];

    \draw[->] (P) -- +(1.7, 1) node[xshift=10, yshift=0] {$V_\rmP$};
    \draw[->] (E) -- +(1.5, 0.4) node[xshift=10, yshift=0] {$V_\rmE$};

    \draw[-, dashed] (P) node[xshift=-25, yshift=0] {$(d_\rmP,h_\rmP )$} -- +(3.5, 0);
    \draw[-, dashed] (E) node[xshift=-25, yshift=0] {$(d_\rmE,h_\rmE )$} -- +(1.2, 0);

    \draw [->] (1,0) arc [radius=1, start angle=0, end angle= 32] node[xshift=15, yshift=-5] {$\gamma_\rmP$};

    \draw [->] (4,4) arc [radius=1, start angle=0, end angle= 15] node[xshift=15, yshift=-5] {$\gamma_\rmE$};
    
    \draw [->] (3,0) arc [radius=3, start angle=0, end angle= 53] node[xshift=17, yshift=-5] {$\beta$};

    \node at (1.8,3) (R) {$R$};  
    
    \end{tikzpicture}
    }
    \caption{ Interception geometry for the pursuer and the evader.    
    }
    \label{fig:interception_geometry}
\end{figure}
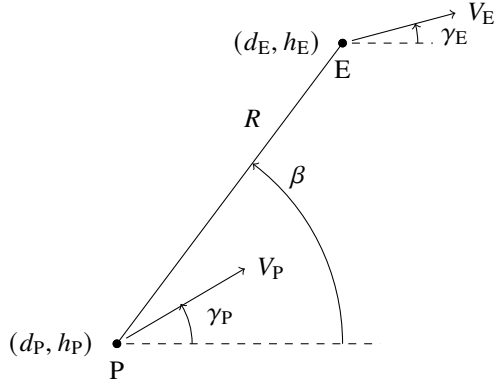

This appendix describes the derivation of the normal acceleration command using proportional guidance. 
A planar interception geometry for a pursuer $\rmP$ and an evader $\rmE$ is shown in Figure \ref{fig:interception_geometry} \cite{kabamba2014fundamentals}.
The interception dynamics for the pursuer and the evader is 
\begin{align}
    \dot V_\rmP 
        &=
            \frac{T_\rmP-D_\rmP}{m_\rmP} - g \sin \gamma_\rmP, \\
    \dot V_\rmE 
        &=
            \frac{T_\rmE-D_\rmE}{m_\rmE} - g \sin \gamma_\rmE, \\
    \dot \gamma_\rmP 
        &=
            -\frac{1}{V_\rmP} \left( \frac{T_{z,\rmP}}{m_\rmP} + g \cos \gamma_\rmP\right), 
    \label{eq:dot_gamma_P}        
    \\
    \dot \gamma_\rmE
        &=
            -\frac{1}{V_\rmE} \left(n_{z,\rmE} + g \cos \gamma_\rmE \right),
\end{align}
where $V_\rmP, V_\rmE$ are the velocities, 
$\gamma_\rmP,$ $\gamma_\rmE$ are the flight-path angles, 
$T_\rmP, T_\rmE$ are the thrust, 
$D_\rmP, D_\rmE$ are the drag, and 
$n_{z,\rmP}, n_{z,\rmE}$ are the normal accelerations of the pursuer and evader, respectively.
The line of sight angle $\beta$ satisfies
\begin{align}
    \dot \beta
        =
            \frac{1}{R}
            \left( 
                V_\rmP \sin(\beta - \gamma_\rmP)
                -
                V_\rmE \sin(\beta - \gamma_\rmE)
            \right)
    \label{eq:dotbeta}
\end{align}

The proportional guidance law is 
\begin{align}
    \dot \gamma_\rmP 
        = 
            \lambda \dot \beta.
    \label{eq:prop_guid}
\end{align}
Substituting \eqref{eq:dot_gamma_P} and \eqref{eq:dotbeta} in \eqref{eq:prop_guid}  yields the normal acceleration command 
\begin{align}
    n_{z,\rmP} 
        &=
            -\frac{V_\rmP}{R}
            \left( 
                V_\rmP \sin(\beta - \gamma_\rmP)
                -
                V_\rmE \sin(\beta - \gamma_\rmE)
            \right)
            -
            g \cos \gamma_\rmP.
\end{align}

\end{document}